# MULTISCALE COMPUTATIONS ON NEURAL NETWORKS: FROM THE INDIVIDUAL NEURON INTERACTIONS TO THE MACROSCOPIC-LEVEL ANALYSIS

**Konstantinos G. Spiliotis and Constantinos I. Siettos***

School of Applied Mathematics & Physical Sciences
National Technical University of Athens
Athens, GR-157 80, Greece

**Abstract.** *We show how the "Equation-Free" approach for multi-scale computations can be exploited to systematically study the dynamics of neural interactions on a random regular connected graph under a pairwise representation perspective. Using an individual-based microscopic simulator as a black box coarse-grained timestepper and with the aid of simulated annealing we compute the coarse-grained equilibrium bifurcation diagram and analyze the stability of the stationary states sidestepping the necessity of obtaining explicit closures at the macroscopic level. We also exploit the scheme to perform a rare-events analysis by estimating an effective Fokker-Planck describing the evolving probability density function of the corresponding coarse-grained observables.*

## 1. Introduction

Over the past years there has been a rapid growth in biological knowledge even at the molecular/cell level concerning the physiology of neurons. Clinical studies and mathematical models have gone hand-in-hand enhancing our understanding and leading to breakthroughs in the field of neuroscience. One of the very first attempts in trying to systematically quantify the dynamics of nerve cells in mathematical terms can be traced back to 1907 when Lapicque introduced the modelling of a neuron's activity by an integrate and fire process using an electric circuit with a capacitor and resistor in parallel driven by a time-varying current (for a discussion see also Abbot, 1999). McCulloch and Pitts in 1943 introduced the concept of the artificial neuron while in 1952 Hodgkin and Huxley proposed the most celebrated model in the field describing the dynamics of action potentials. For their work in the area the two biophysicists won the Noble Prize in Physiology and Medicine in 1963. Other early benchmarks that shaped the field include the work of Rall [1959] who actually originated the development of compartmental models which incorporate spatial electro-physiological characteristics, the spatial distributed Fitzhugh-Nagumo model [Fithugh, 1961; Nagumo et al., 1962] describing the propagation of nerve pulses in the form of traveling waves, the work of Wiener and Rosenbluth [1946] and Greenberg and Hastings Cellular Automata [1978] modeling complex spatial patterns in excitable media with interconnected neurons, and models of spatially localized neural populations [Smith & Davidson, 1962; Griffith, 1963; Anninos et al., 1970; Wilson & Cowan, 1972]. A review on recent developments on the modelling of neuron dynamics can be found in [Inzikievitz, 2004; Herz et al., 2006].

The studies have proceeded to the development of detailed state-of-the-art models aspiring to approximate the complex dynamics underlying the physics and mechanisms of a wide range of problems stretching from the description of the behaviour of certain neural tissues such as the cerebral neocortex [Marr, 1970], the cortical and thalamic [Wilson & Cowan, 1973], the CA3 hippocumpal [Traub, 1983] to visual hallucinations [Ermentrout & Cowan, 1979; Bressloff et al., 2001], phase transitions in human hand movements [Haken et al., 1985] and working memory mechanisms [Durstewitz et al., 2000; Brunel & Wang, 2001; Laing et al., 2002; Laing & Troy, 2003; Durstewitz and Seaman, 2006] and from neurological disorders such as schizophrenia [Cohen et al., 1996; Hoffman, 1997; Rolls et al., 2008] and epilepsy dynamics [Babloyantz & Destexhe, 1986; Theoden et al., 2004] to therapeutic surgical procedures such as deep-brain stimulation process [Rubin & Terman, 2004] to name just a few.

These models are usually individual-based ones: they are composed by a large number of individual subunits interacting through a network, a caricature of the connections circuitry [Strogatz, 2001]. The dynamics of each single subunit are described in the spirit of representations such as the Hodgkin-Huxley [Gutkin et al., 2001; Laing & Chow, 2001; Laing & Chow, 2002] or Fitzhugh-Nagumo equations [Coombes & Osbaldestin, 2000], the integrate and fire mechanism [Schuster & Niebur, 1993; Brunel & Hakim, 1999; Bressloff & Coombes, 1999; Omurtag et al., 2000; Casti et al., 2002; Ermentrout & Chow, 2002] phase and nonlinear oscillators [Kuramoto, 1991, Ermetrout, 1992; Abbott & van Vreeswijk, 1993; Murray, 2007], cellular-automata-like spiking propagation [Wolfram, 1984; Kaneko, 1991; Kozma et al., 2005; Furtado & Copelli, 2006]. A review of such models can be found in [Borisyuk et al., 2002].

*Corresponding author; e-mail: ksiet@mail.ntua.gr

Due to the strongly heterogeneous nature of the nerve cells as well as their stochastic and nonlinear very-large scale interactions, the collective –macroscopic - behaviour of such multicellular systems is far from intuitive [Destexhe & Contreras, 2006] and cannot, most of the time, be straightforward predicted. These large-scale subunits interactions result to a remarkable and fundamental feature of analogous systems: the collective–emergent-macroscopic – behaviour is far from intuitive [and cannot, most of the time, be straightforward predicted. Sustained oscillations, travelling waves, multiplicity of stationary states and spatio-temporal chaos [Kopel & Ermetrout, 1986; Kaneko, 1992; Chow, 1998; Freeman, 1999; Bressloff & Coombes, 1999; Chow & Koppel, 2000; Coombes & Osbaldestin, 2000; Haskell et al., 2001; Roxin et al., 2004; Coombes, 2005] are some examples of the rich nonlinear dynamics emerging in the coarse-grained population-level.

For the systematic analysis of such behaviours, one of the most critical issues in the area is related to the bridging of the different scales: the micro-scale, where the interactions of the subunits take place and the macroscopic scale where the dynamic behaviour of the constitutive system/ tissue/organ/ emerges. Traditionally, the gap between high-dimension (microscopic) and low-order (coarse-grained level) is bridged through closures, relating higher-order, fast moments of the evolving detailed distributions to a few, low-order, slow, "master" moments of the underlying detailed distributions [see for example: Treves, 1993; Bressloff et al., 1997; Bressloff, 1999; Haskell et al., 2001; Casti et al., 2002; Cai et al., 2004; Rangan & Cai, 2006]. However no general systematic methodology for deriving such closures exists; they are often based on *ad hoc* assumptions such as identical subunits, completely regular connections, and infinite populations, that may introduce a systematic bias and possibly miss important qualitative at the coarse-grained level [see for example the critical discussion on the influence of certain closure approximations applied to integrate and fire networks in [Ly& Tranchina, 2007].

Good closures in the form of ordinary, or partial differential equations (ODEs or PDEs), describing the coarse-grained dynamics may, in principle, be extracted from the individualistic interactions, but usually are unavailable-or extremely difficult to derive. The problem is that due to heterogeneities, nonlinearities and stochastic interactions serious limitations come up when trying to systematically bridge the different time and spatial scales, in order to extract effective-macroscopic equations.

Without the availability of such good closures, what is usually done is the exploration of the coarse-grained dynamics by performing long runs in time using the available microscopic simulator starting from different initial distributions and averaging over many ensembles. However this approach is both computational consuming and inadequate for important tasks such as the systematic detection of coarse-grained criticalities, stability analysis and the computation of coarse-grained bifurcation diagrams. There is an arsenal of state of the art computational tools that are good for such system-level analysis, yet these rely on the availability of closed form coarse-grained models.

In this work we show how the so-called "Equation-free" approach [Gear et al., 2002; Makeev et al., 2002; Runborg et al., 2002; Kevrekidis et al., 2003; Siettos et al., 2003; Kevrekidis et al., 2004; Haataja et al., 2004; Mooller et al., 2005; Moon et al., 2005; Russo et al., 2007; Laing & Kevrekidis, 2008] a computational framework for multi-scale computations can be used to efficiently extract "systems-level" information from detailed-individualistic models describing the dynamic behaviour of neurons interacting on a network under a pairwise representation perspective [Kelling & Eames, 2005; Schneidman et al., 2006].

For our illustrations we used a simple dynamic individual--based model based on the work presented in [Kozma et al., 2005]. Here, each neuron can be in one of two states: excited (activated) or non-excited (inactivated). Neurons interact with their neighbors on a random regular network [Bollobás, 2001; Newman, 2003] in a bidirectional way and change their states over discrete time in a probabilistic manner according to simple rules involving their own states and the states of their links. The relatively simple rules governing the interactions of the subunits result to the emergence of complex dynamics in the coarse-grained level including phase transitions from low to high density of excited neurons and multiplicity of coarse-grained stationary states.

The Equation-Free approach is exploited to perform system-level tasks such as bifurcation, stability analysis and estimation of mean appearance times of rare events, circumventing the need for obtaining coarse-grained models in closed form. Using the microscopic simulator as a black-box timestepper of the network excitation density we compute the coarse-grained equilibrium bifurcation diagrams and examine the stability of the solution branches with respect to a parameter describing how the state of a neuron affects the state of its links. Coupling the Equation-Free approach with Kramers' theory [Hänggi et al., 1990; Haataja et al., 2004; Sriraman et al., 2005] we also compute the mean transitions rates between the apparent coarse-grained equilibria. The required closures are obtained by relating higher-order moments of the state distribution with lower-order ones "on demand' combining the Equation-Free approach with Simulating Annealing.

The paper is organized as follows: in section 2.1 we give a short description of the basic concept of the Equation-Free framework; in section 2.2 we describe how the Equation-Free approach combined with Simulated Annealing can be exploited to enable microscopic simulations to converge on the coarse-grained slow manifold, while in section 2.3 we show how it can be combined with Kramers' theory for rare-event calculations. In section 3 we describe the individual-based model and in section 4 we present and discuss the simulation results; finally we conclude in section 5.

## 2. The Equation-Free approach for multi-scale computations

### 2.1 The basic concept

Consider a complex system for which the only good modelling description is available on the microscopic (atomistic, molecular, agent-based) level. We also believe that macroscopic laws in the form of ordinary and/or partial-integro-differential equations (ODEs/PDEs) which can describe the coarse-grained dynamics of the underlying physics may be in principle derived but due to the inherent complexity of the problem, accurate quantitative closures have not yet been extracted.

For the sake of presentation, lets us write these macroscopic equations in the form of

$$\frac{d\,\boldsymbol{u}_s}{dt} = \boldsymbol{F}(\boldsymbol{u}_s, \boldsymbol{p}), \tag{1}$$

where $\boldsymbol{u}_s \in R^n$ denotes a low-dimensional coarse-grained state vector, observable through measurement, $\boldsymbol{p} \in R^m$ is a vector denoting the system parameters, and $\boldsymbol{F}: R^n \times R^m \to R^n$ is a not explicitly available operator. The question is how to systematically perform system-level tasks (e.g. construct bifurcation diagrams of steady and time-dependent solutions and perform their stability analysis) when direct simulation of the microscopic code is the only possible choice and despite the fact that $\boldsymbol{F}$ is not explicitly available.

The answer comes from the concept of the coarse timestepping [Gear et al., 2002; , Makeev et al., 2002; Runborg et al., 2002; Kevrekidis et al., 2003; Siettos et al., 2003; Kevrekidis et al., 2004; Haataja et al., 2004; Möller et al., 2005].

A brief schematic of the concept of the coarse timestepper is given in Figure 1. The coarse timestepper is a fundamental building block in the Equation-Free framework. It is the way to efficiently obtain macroscopic input–output information from a microscopic simulator bypassing the need to extract macroscopic equations in a closed form. Let $U(x)$ denote the distribution function of the microscopic variables over the set of microscopic coordinates x. The main assumption behind the coarse timestepper is that a coarse-grained model for the fine-scale dynamics exists and closes in terms of a few coarse-grained variables, say, $\boldsymbol{u}_s$. Typically these are usually the low-order moments of the microscopically evolving distributions. The existence of a coarse-grained model implies that the higher order moments, say, $\boldsymbol{u}_f$, of the distribution $U(x)$ become, relatively quickly over the time scales of interest, "slaved" to the few lower ones $\boldsymbol{u}_s$. This scale separation could be viewed in the form of a singularly perturbed system, reading:

$$\frac{d\,\boldsymbol{u}_s}{dt} = \boldsymbol{h}_s(\boldsymbol{u}_s, \boldsymbol{u}_f, p) \tag{2a}$$

$$\varepsilon \frac{d\,\boldsymbol{u}_f}{dt} = \boldsymbol{h}_f(\boldsymbol{u}, \boldsymbol{u}_f, \boldsymbol{\mu}) \tag{2b}$$

with $\varepsilon \ll 1$. Eq. (2a) corresponds to the "slow" dynamics and Eq. (2b) to the "fast" ones. Notice that under certain conditions [Fenichel, 1979] the overall dynamics defined by Eq. (2) will approximate very quickly the dynamics of the following system

$$\frac{d\,\boldsymbol{u}_s}{dt} = \boldsymbol{h}_s(\boldsymbol{u}_s, \boldsymbol{u}_f, \boldsymbol{\mu}) \tag{3a}$$

$$0 = \boldsymbol{h}_f(\boldsymbol{u}, \boldsymbol{u}_f, \boldsymbol{\mu}) \tag{3b}$$

The above system can-under some mild assumptions [Fenichel, 1979; Gear et al., 2005]- be written in the form of Eq. (1) by applying the implicit function theorem for eq. (3b), i.e.,

$$\boldsymbol{u}_f = q(\boldsymbol{u}_s, \boldsymbol{\mu}), \quad \boldsymbol{F}(\boldsymbol{u}_s, \boldsymbol{\mu}) \equiv \boldsymbol{h}_s(\boldsymbol{u}_s, q(\boldsymbol{u}_s, \boldsymbol{\mu}), \boldsymbol{\mu}) \tag{3c}$$

where $q$ is a continuous function defining the relation between the slow and the fast variables after a short time horizon, i.e. the slow manifold on which the dynamics of the system evolve after a fast transient phase (see figure 2).

What a coarse time-stepper does, in fact, is providing such closures on demand (relatively short bursts of fine scale simulation naturally establish this slaving relation (refer to [Gear et al., 2002; Makeev et al., 2002; Runborg et al., 2002; Kevrekidis et al., 2003; Siettos et al., 2003; Kevrekidis et al., 2004; Gear et al., 2005] for more detailed discussions). Briefly, once the appropriate macroscopic observables have been identified, the coarse timestepper consists of the following essential components:

(a) Prescribe a coarse-grained initial condition (e.g. the density of activated neurons in the network) $u(t_0)$.
(b) Transform it through a lifting operator $\mu$, to consistent microscopic realizations: $U(t_0) = \mu\, u(t_0)$
(c) Evolve these realizations in time using the microscopic simulator for the desired short macroscopic time $T$, generating the value(s) $U(t_0 + T)$. The choice of $T$ is associated with the (estimated) spectral gap of the linearization of the unavailable closed macroscopic equations [Kevrekidis et al., 2003; Siettos et al., 2003].
(d) Obtain the coarse-grained variables using a restriction operator $M$: $u(t_0 + T) = M U(t_0 + T)$.

The above procedure can be considered as an "black box" coarse timestepper that, given the initial state of the system $(u(t_0), \mu)$ reports the solution, the result of the integration, after a given time horizon $T$, i.e.,

$$u(t_0 + T) = \Phi_T (u(t_0), \mu), \qquad (4)$$

where $\Phi_T$ is the temporal evolution operator of the system. At this point one can:

(e) Implement a computational superstructure such as Newton's method [Kelley, 1995] or Arnoldi's method [Saad, 1992; Christodoulou and Scriven, 1998] or Newton- Picard method (e.g the Recursive Projection Method of Shroff and Keller (1993); Kelley et al., 2004) as a shell around the coarse-timestepper to extract systems-level information. For example coarse-grained (macroscopic) steady states can be obtained as fixed points, using $T$ as sampling time, of the mapping $\Phi_T$:

$$u - \Phi_T (u, \mu) = 0 \qquad (5)$$

If on the other hand, a periodic oscillatory behavior is observed then one seeks for solutions which satisfy

$$u(0) = u(T), \qquad (6)$$

with $T$ denoting now the period of oscillation. Periodic solutions can be computed as fixed points of the mapping (5) augmented by the so-called phase constraint (also called a pinning condition)

$$g(u, \mu, T) = 0, \qquad (7)$$

which factors out the infinite members of the family of periodic solutions [Lust et al., 1998; Kavousanakis et al., 2008].

**2.2 Using the Equation-Free approach to converge on the slow manifold**
For the scheme to be accurate, the overall procedure has to be applied when the system evolves on the slow manifold (i.e., the fixed point iteration (3) has to be solved *on* the slow manifold). However, in general, the lifting operator creates microscopic distributions off the slow manifold. If the time required for trajectories emanating from initial conditions off the slow manifold to reach the slow manifold is very small compared to $T$ the above requirement is satisfied for any practical means. But still we can farther enhance our calculations by forcing our system to start from consistent to the coarse-grained variables (lower-order moments of the microscopic distribution) microscopic initial conditions *on* the slow manifold. This can be achieved by exploiting the Equation-Free approach as follows (see figure 3):

(1) Prescribe the desired coarse-grained initial conditions $\boldsymbol{u}_s(t_0)$.

(2) Transform $\boldsymbol{u}_s(t_0)$ through a lifting operator $\boldsymbol{\mu}$ to consistent microscopic realizations: $U(t_0) = \boldsymbol{\mu}\,\boldsymbol{u}(t_0)$. In general the higher order moments of the microscopic distributions, say $\boldsymbol{u}_f[U(t_0)]$, will be off the slow manifold.

*k=0;*
*Do until convergence to the slow manifold {*
  *k=k+1;*
  (3) Evolve these realizations in time using the microscopic simulator for a very short macroscopic time $dT \ll T$, generating the value(s) $U^k(t_0 + dT)$ and compute the higher moments (for practical means, up to a specific order *l*) $\boldsymbol{u}_f[U^k(t_0 + dT)]$.

  (4) Restrict back to the prescribed coarse-grained initial conditions $\boldsymbol{u}_s(t_0)$, preserving the values of $\boldsymbol{u}_f[U^k(t_0 + dT)]$.

At this point Simulated Annealing (SA) [Kirkpatrick et al., 1983; Cerny, 1985; Aarts et al., 2005] can be employed to obtain the desired micro-structure. The objective function at the step j of the SA algorithm may be defined as $E^j = \|\boldsymbol{u}_f^{\,j} - \boldsymbol{u}_f(U^k)\|$, where $\boldsymbol{u}_f(U^k)$ are the target values of the fast variables. In particular for a network of neurons with discrete states (e.g. active/ inactive), the fast variables $\boldsymbol{u}_f$ correspond to the densities of pairs, triples, quadruples links of neurons of certain states in the network which are related to the second, third, and fourth respectively spatial moments of the underlying distribution [Murrell et al., 2004]. In this case the SA algorithm may read as follows:

  *0) Set the initial systems pseudotemperature Temp and select the annealing schedule, i.e. the way the speudotemperature will decrease.*

  *Do until convergence {*
    *1) Evaluate the pseudoenergy (objective function) $E(\boldsymbol{u}_f)$ of the network;*

    *2a) Select randomly a neuron (or set of neurons)*
    *2b) Select randomly another neuron (or set of neurons)*
    *2c) Create a new distribution over the network by switching their states. During switching (e.g. from an active to active neuron and vice versa) the first-order moment does not change. Let the new values of the fast variables corresponding to the new network configuration be given by $\boldsymbol{u}'_f$.*

    *3) Evaluate the new objective function $E(\boldsymbol{u}'_f)$.*

    *4) Accept or reject the new network configuration using the Metropolis procedure [Metropolis, 1953]:*

      *4a) Accept the new configuration if $E(\boldsymbol{u}'_f) < E(\boldsymbol{u}_f)$,*

      *4b) Accept the new configuration if $E(\boldsymbol{u}'_f) > E(\boldsymbol{u}_f)$, with a probability*
      $$\exp\left[-\left(E(\boldsymbol{u}'_f) - E(\boldsymbol{u}_f)\right)\Big/Temp\right];$$
      *otherwise reject it.*
    *6) Reduce the system pseudotemperature according to the annealing schedule.*

  *} End Do*

*} End Do*

## 2.3 Rare-events analysis
Coupling the framework with Kramers' theory [Kramers, 1940; Hänggi et al., 1990; Haataja et al., 2004;

Sriraman et al., 2005; Kopelevich et al., 2005; Erban et al., 2006] we can also compute the mean transitions rates between the apparent coarse-grained equilibria.

The coarse timestepper is used to approximate from short runs the drift $u(\psi)$ and from the time dependence of the variance the diffusion coefficient $D$ of a Fokker-Planck equation of the form

$$\frac{\partial P(\psi,t)}{\partial t} = \left[ -\frac{\partial}{\partial \psi} u(\psi) + \frac{\partial^2}{\partial \psi^2} D(\psi) \right] P(\psi) \tag{8}$$

where $P(\psi,t)$ is the probability density function of the coarse-grained coordinate $\psi$, $u(\psi)$ is the drift and $D(\psi)$ is the diffusion coefficient.

The drift and the diffusion coefficients can be estimated by appropriate initialized short-time simulations using the coarse timestepper from the following relations:

$$u(\psi) = \lim_{\Delta T \to 0} \frac{\langle \Delta \psi(t,\psi) \rangle}{\Delta T}, \qquad 2D(\psi) = \lim_{\Delta T \to 0} \frac{\langle \Delta \psi(t,\psi)^2 \rangle}{\Delta T}, \qquad \text{where}$$

$\Delta \psi(t,\psi) = \psi(t+\Delta T,\psi) - \psi(t,\psi)$.

When the potential barrier between the stable and the unstable point is relatively big to the noise, the mean escape rates between two coarse-grained equilibria can be estimated by the formula [Erban et al., 2006]:

$$\tau \approx \int_{\psi_{stable}}^{\psi_{unstable}} exp[\beta G(\psi)] d\psi \int_{-\infty}^{\psi_{unstable}} \frac{1}{D(\psi)} exp[-\beta G(\psi)] d\psi \tag{9}$$

where $\beta G(\psi)$ stands for free-energy of the system, which is a function of the drift and the diffusion coefficient given by:

$$\beta G(\psi) = -\int_0^{\psi} \frac{u(\psi')}{D(\psi')} d\psi' + \ln D(\psi) + const \tag{10}$$

$\psi_{stable}$ and $\psi_{unstable}$ are the values of the reaction coordinate $\psi$ at the minimum (corresponding to the stable point) and at the maximum (corresponding to the unstable point) of $\beta G(\psi)$

### 3. The atomistic-based model

For our illustrations the atomistic model considered here is based on the theory of neuropercolation [Kozma et al., 2005]. The model consists of $N$ neurons that interact through a random regular connected graph [Bollobás, 2001; Newman, 2003] defining the connections between these neurons. The connectivity degree $d$ between neurons is considered to be a constant. The links are bidirectional and fixed during the simulations and loops (connections of individuals to themselves) are not allowed.

Each neuron is labeled by an index $i = 1,2,...N$ and is characterized by two states: the state "1" if is it activated and the state "0" if it is inactivated. At each discrete time step $t$ neurons interact with their links and change their states in a probabilistic manner according to following simple rules involving their own states and the states of their links:

- An inactivated neuron $i$ becomes activated with a probability $p_{0 \to 1} = \varepsilon$ if the number of activated links of the neuron (including the state of neuron $i$) is less than $(d+1)/2$ *and at least one of each links is activated*; otherwise the probability to get activated is set to $p_{0 \to 1} = 1 - \varepsilon$.

- An activated neuron becomes inactivated with a probability $p_{1 \to 0} = 1 - \varepsilon$ if the number of activated links of the neuron (including the state of neuron $i$) is less than $(d+1)/2$; otherwise the probability is set to $p_{1 \to -0} = \varepsilon$.

where $\varepsilon \in (0 \quad 0.5)$.

In Figure 4 we give specific examples illustrating the functionality of the above rules.

## 4. Simulations results and discussion

The simulation results were obtained using a total of $N = 20000$ neurons, while $d$ is set equal to four. The random regular graph was generated using the procedure described in [Steger & Wormald, 1999]: an unlinked pair $(i, j)$ of individuals with $i \neq j$ is randomly chosen from a total of $Nd$ even number of points ($d$ points in $N$ groups); connect $i$ with $j$ setting $G_{ij} = 1$; leave $G_{ij} = 0$ otherwise; repeat the above procedure until all individuals are properly linked and the graph is complete, i.e. the infection can potentially reach every individual in the network when starting from any other one.

Figure 5 depicts the time evolution of the density of activated neurons, say $p$, for various values of the activation probability $\varepsilon$ (the density of inactivated neurons, is given by $p_c = 1 - p$). As it is clearly seen, the model exhibits some interesting nonlinear dynamics. Specifically, for small values of the parameter ε and depending on the initial conditions there are two stable stationary states one corresponding to the "all-off" state (all neurons turn to be inactivated) and the other to a partially activated network (see figure 5a). As the value of $\varepsilon$ increases - and now independently on the initial conditions - the second stable stationary point disappears and the network converges to the "all–off" state which is the only possible one (figure 5b). The "all-off" state looses its stability for bigger values of $\varepsilon$ making the network to converge to a single non-zero stable stationary state (figure 5c, d). The above results are indicative of the existence of several critical parameter values marking the onset of qualitative transitions in the dynamic behavior of our system - and as the bifurcation theory suggests - the emergence of unstable coarse-grained states which are unreachable through plain long-run temporal simulations.

At this point it is interesting to note that in the case of the two stable stationary states the non-zero is a metastable one: due to stochasticity the network's state is prone to jump to the "all-off" region of attraction (see figure 6). Based on the above simulation results, there are three obvious questions to be answered here: (a) which is the complete set of fixed point solutions (both stable and unstable ones), (b) which are the critical points which mark the onset of the phase transitions, and, (c) what is the transition rate of a stable metastable state emerging for small values of ε. We answer all three questions by exploiting the Equation-Free approach. The first two questions questions are being answered by performing coarse-grained bifurcation analysis while the third one by exploiting Kramers' theory for rare-events calculations.

As described in section 2.1 the first step in the Equation-Free framework is the selection of the coarse-grained statistics. The assumption is that a good closure in principle exists in terms of a set of a few coarse-grained observables which correspond to some lower order moments of the microscopic distribution. This means that the long-term dynamic behavior of the problem lies on a low-dimensional manifold parametrized by these few coarse-grained observables. Higher-order moments become quickly, relative to the macroscopic time, functions of the lower-order ones. But in what level does this happen? In order to give an answer to this question we constructed the phase portrait of the density of the activated neurons, which corresponds to the first order moment of the microscopic distribution and the densities of pairs between the states of the neurons, which correspond to the second spatial moment of the network distribution [Murrell et al., 2004].

These are the densities of the activated neurons linked to activated neurons ($\{11\}$ pairs), the activated neurons linked to inactivated neurons ($\{10\}$ pairs) and the inactivated neurons linked to inactivated neurons ($\{00\}$ pairs). Under this pairwise representation, the density of $\{01\}$ pairs is equal to the $\{10\}$ one. Figure 7 depicts the phase portraits of the $p$ vs $\{10\}$ pair density around a nonzero stationary state at $\varepsilon = 0.14$. The trajectories were obtained by averaging over 500 ensemble realizations. We were able to create microscopic state configurations consistent with the first, second and third (which correspond to triples) -order moments by utilizing the SA procedure, thus initializing these coarse-grained quantities at will. As it is clearly shown after an initial short transient that the dynamics are evolving on a one-dimensional slow manifold, parametrized by the first-order moment- towards the coarse-grained stable stationary state; the higher order moments become functions of the first one. This implies that the expected dynamics over the ensembles conditioned on this slow manifold through SA closes and hence can be described by just the first-order moment, i.e. the density of activated (or inactivated) neurons.

Hence it makes sense to search for coarse-grained equilibria as fixed points of the coarse timestepper

$$p = \Phi_T(p, \varepsilon) \tag{12}$$

the result of lifting consistently an initial distribution conditioned on the slow manifold using SA with $dT = 1$ time step and evolving on the slow manifold for a time interval of $T = 5$ *time steps* to the final distribution restricted back to the first-order moment by averaging over 10000 realizations. The coarse-grained derivatives for the fixed point iteration (here, Newton-Raphson) were estimated numerically through the coarse timestepper, by appropriately, perturbing the coarse-grained initial conditions, constrained again with SA on the slow-manifold.

The fixed-point solutions of (12) were obtained using the coarse timestepper-based fixed point algorithm over 10000 consistent to the coarse-grained variable *p* microscopic network realizations, augmented by the pseudo arc-length continuation [Chan & Keller, 1982]:

$$G(p,\varepsilon) = p - \Phi_T(p,\varepsilon) = 0 \tag{13a}$$

$$N(p,\varepsilon) = \frac{dp_1}{ds}(p - p_1) + \frac{d\varepsilon_1}{ds}(\varepsilon - \varepsilon_1) - \Delta s = 0 \tag{13b}$$

where $(p_1, \varepsilon)$ and, $(p_0, \varepsilon_0)$ are two previously calculated solutions and $\Delta s$ is the pseudo arc-length continuation step.

The procedure involves the iterative solution of the following linearized system

$$\boldsymbol{J}^k \begin{bmatrix} \delta p^{k+1} - \delta p^k \\ \delta \varepsilon^{k+1} - \delta \varepsilon^k \end{bmatrix} = -\begin{bmatrix} G(p,\varepsilon) \\ N(p,\varepsilon) \end{bmatrix}^k$$

where *k* is the iteration counter and $(p,\varepsilon)^k$ is the approximation of the solution at the *kth* iteration. $\boldsymbol{J}^k$ is the jacobian matrix:

$$\boldsymbol{J}^k \equiv \begin{bmatrix} 1 - \frac{\partial \Phi_T}{\partial p} & \frac{\partial \varepsilon}{\partial p} \\ \frac{dp_1}{ds} & \frac{d\varepsilon_1}{ds} \end{bmatrix}^k$$

which can be approximated numerically by calling the timestepper at appropriately perturbed values of the corresponding unknowns. In our difference approximation we used perturbations of the order of $10^{-2}$. The above computational framework enables the microscopic simulator to converge, to its own coarse-grained stable and unstable (which are not reachable through long run temporal simulations) points and trace their locations with respect to the bifurcation parameter. The eigenvalues (here is just one) of the Jacobian $\frac{\partial \Phi_T}{\partial p}$ determine the local stability of the stationary solutions: a fixed point is stable when the modulus all eigenvalues is smaller than one and unstable if there exists at least one eigenvalue with modulus greater than one.

Figure 8 shows the derived coarse-grained bifurcation diagram of *p* vs *ε*; solid lines correspond to stable stationary states while dotted lines to unstable ones. The dependence of the one and only, in our case, eigenvalue with respect to the bifurcation parameter *ε* is shown in the inset of figure 8.

For small values of *ε* the solutions bifurcate through a turning point around *ε* = 0.165 which marks the change in stability of the non-zero stationary network phase. The "all-off" phase, which exists for all values of *ε*, is stable for small values of *ε* and unstable beyond a transcritical point around *ε* = 0.22 giving rise to a branch of a stable non-zero phase of relatively low network activities.

As demonstrated already in figure 6, the relatively high activity stable phase is metastable: after a relatively long period of time, depending on the value of *ε* (the closer to the turning point the smaller and vice versa) the network can in principle "jump" from a high activation to the "all-off" phase. The mean transition rates for a particular value of our bifurcation parameter *ε* was calculated in a systematic, computational efficient way by approximating the coarse-grained free-energy of the network, as described in section three.

To this end, figure 9 depicts the free energy as computed by short runs and 100000 realizations at *ε* = 0.162. Here the reaction coordinate is the distance from the density of activated neurons *p* at the node, i.e. $\psi = p - p_{node}$. Again in order to calculate the drift and the diffusion coefficient as they are defined in (9), we used the procedure of section 2.3: for each $\psi$ we lifted to consistent - to the coarse-grained variable *p* - microscopic network realizations whose second-order moments (the densities of pairs) lie on the slow manifold. Using Eq. (10) we found a mean first-passage time of the order of 20000 time steps which is in good agreement with the outcomes of the long-run temporal simulations.

## 5. Conclusions

We have demonstrated how the Equation-Free approach, a computational framework for multi-scale computations can be exploited to systematically analyse the coarse-grained dynamics of a network of neurons. For our illustrations we used a extremely simple individual-based model which, however, is able to catch a fundamental feature of such problems which is the emergence of complex dynamics in the coarse-grained level including phase transitions from high to low (or even to zero) density of excited neurons and multiplicity of coarse-grained stationary states. We used the concept of the coarse timestepper and that of Simulated Annealing to enable the microscopic simulator to converge to its own coarse-grained slow manifold on which the coarse-grained dynamics evolve after a fast-in the macroscopic scale- time interval. Using the scheme we were able to track the coarse-grained equilibria (stable and unstable) and examine their stability with respect to a parameter representing the neurons excitation probability. We detected the coarse-grained critical points marking the onset of the phase transitions and we derived a Fokker-Planck equation to approximate the mean escape rates of the metastable coarse-grained equilibria. Further research could proceed along several directions including the systematic analysis of models closer to the real-physics, the use of more realistic network structures, advanced techniques such as the Computational Singular Perturbation method for the systematic reconstruction of the slow manifold [Lam & Goussis, 1994; Goussis & Valorani, 2006] as well as the use of Diffusion Map techniques [Coifman et al., 2005; Kolpas et al., 2007] that can be exploited to extract the right coarse-grained variables when dealing with more complex physics.

**Acknowledgments**
This work has been funded by the project PENED 2003. The project is cofinanced 75% of public expenditure through EC-European Social Fund, 25% of public expenditure through Ministry of Development - General Secretariat of Research and Technology and through private sector, under measure 8.3 of OPERATIONAL PROGRAM "COMPETITIVENESS" in the 3rd Community Support Program.

# Figures

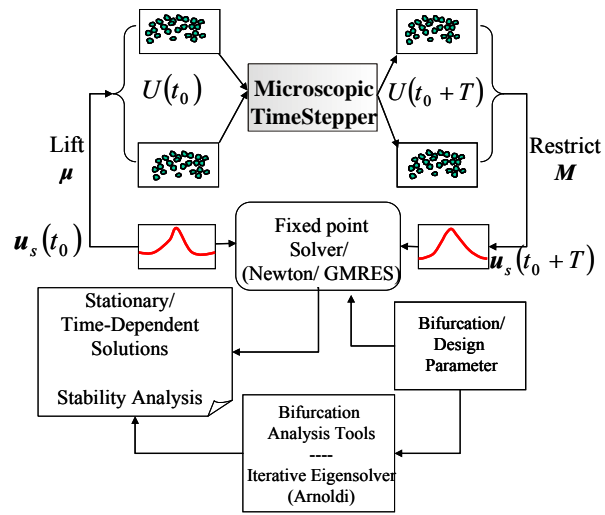

**Figure 1**. Schematic of the Coarse-Timestepper.

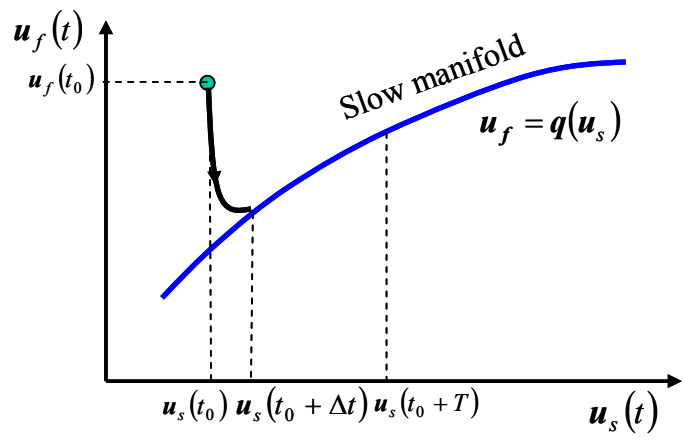

**Figure 2**.

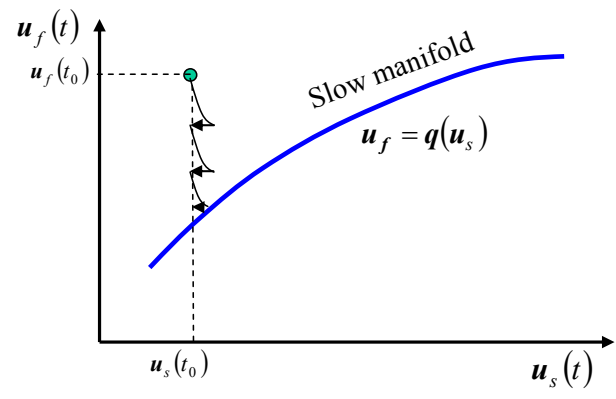

**Figure 3**.

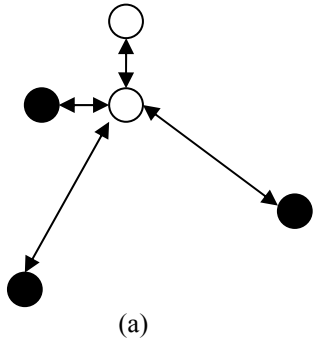 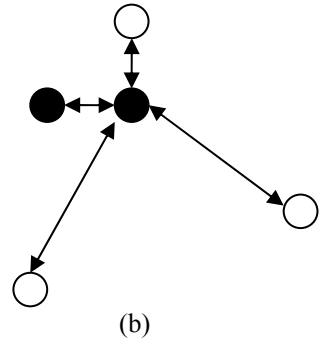

**Figure 4**.

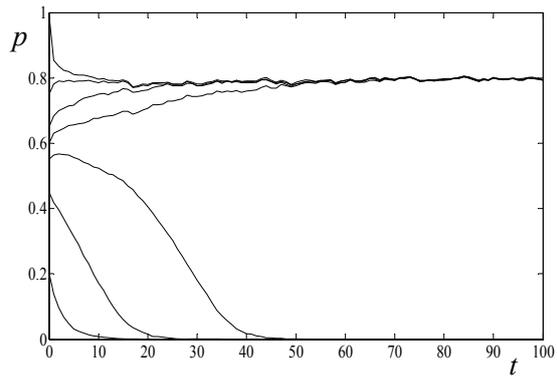
(a)

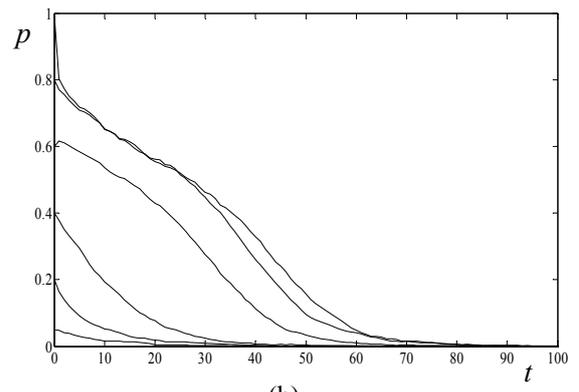
(b)

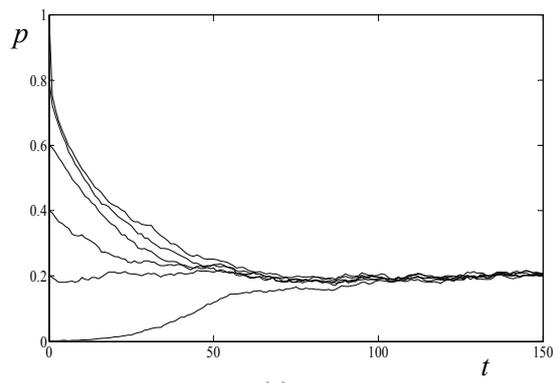
(c)

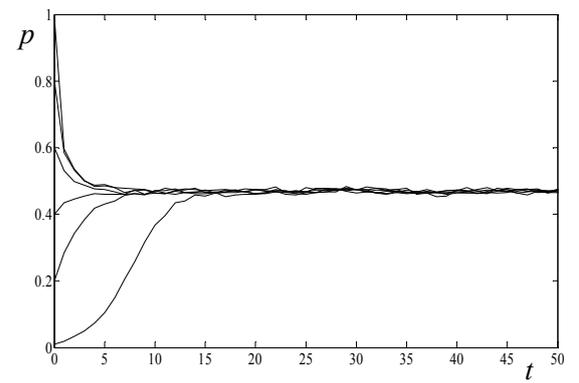
(d)

**Figure 5**.

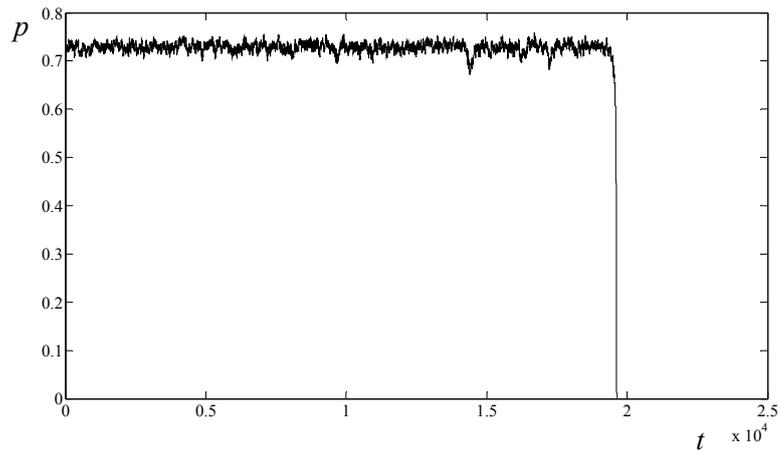

**Figure 6.**

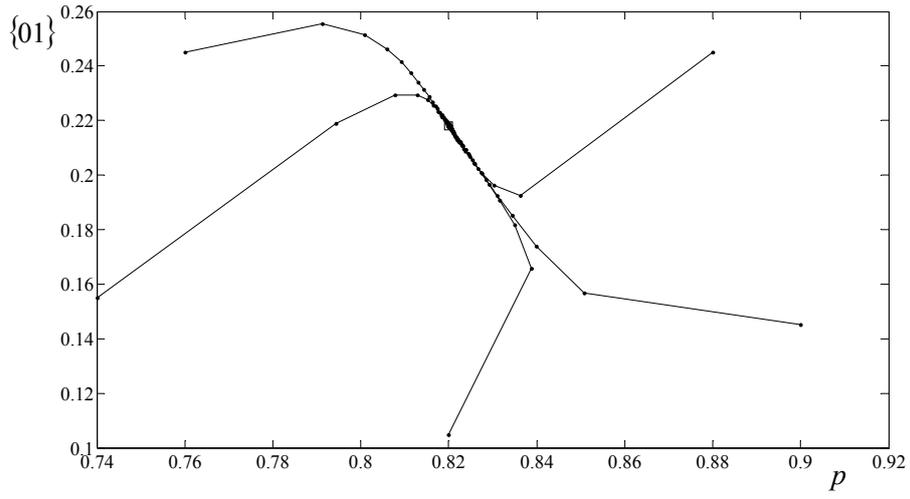

**Figure 7**.

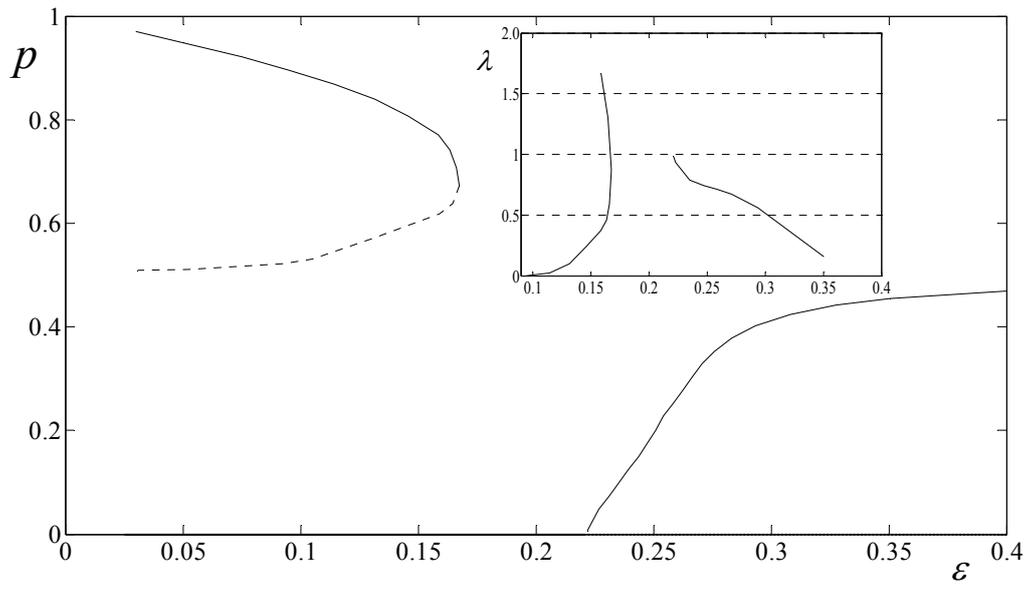

**Figure 8**.

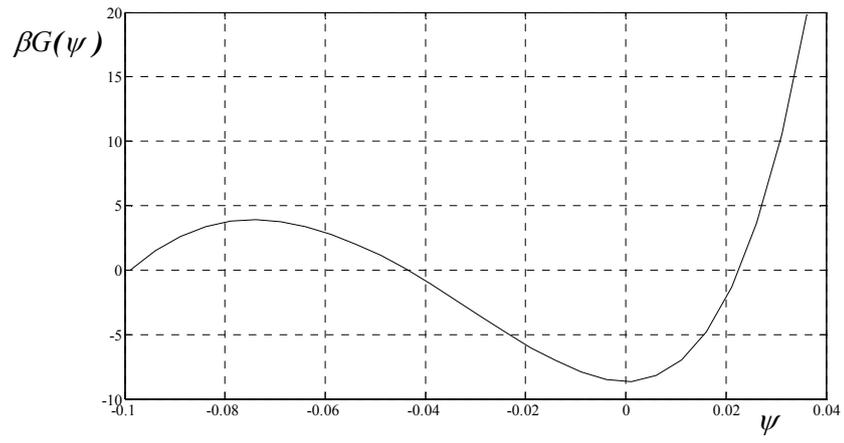

**Figure 9**.

# Captions of Figures

**Figure 1**. Schematic of the Coarse-Timestepper.

**Figure 2**. The fundamental assumption of the Equation-Free approach: Higher order moments ($u_f$) of an evolving microscopic distribution become "quickly" within the time-horizon $T$ functionals to the lower-order moments ($u_s$).

**Figure 3**. Using the Equation-Free approach to successively converge to the attracting slow manifold starting from coarse-grained initial conditions off the manifold. Simulated annealing is employed to obtain the desired micro-structure at each step.

**Figure 4**. Examples of the evolution rules. Filled (Open) circles represent activated (inactivated) neurons: (a) $p_{0 \to 1} = 1 - \varepsilon$, since there are 3 activated neurons out of 5 links, (b) $p_{1 \to 0} = 1 - \varepsilon$, since there are 2 activated neurons out of 5 links.

**Figure 5**. Temporal simulations illustrating the evolution of the density of activated neurons starting from different initial conditions as the value of the parameter $\varepsilon$ increases: (a) $\varepsilon = 0.15$, (b) $\varepsilon = 0.20$, (c) $\varepsilon = 0.25$, (d) $\varepsilon = 0.4$. The appearance of two stable stationary solutions is possible for small values of $\varepsilon$.

**Figure 6.** A long-run temporal evolution of one microscopic realization of the density of activated neurons for $\varepsilon = 0.163$.

**Figure 7**. Phase portraits of $p$ vs {10} pair density for $\varepsilon = 0.14$; the trajectories were averaged over 500 copies; the stationary state is marked with the square marker. The initial conditions corresponding to different densities of pairs on the graph were created by simulated annealing. The dynamics are characterized by two stages: a fast approach (within around 3 time steps) to a one dimensional slow manifold parametrized by the first-order moment and then a slower decay to the stationary state (marked with the square) on the "slow manifold" on which all long-term dynamics evolve.

**Figure 8**. The one-dimensional coarse-grained bifurcation diagram of the density of activated neurons $p$ (first-order moment) with respect to model parameter $\varepsilon$. Solid lines correspond to stable coarse-grained equilibria, while, the dotted line corresponds to unstable ones. The inset depicts the computed eigenvalue determining the systems stability.

**Figure 9**. The coarse-grained free energy for $\varepsilon = 0.163$ with respect to the distance, $\psi$ from the metastable point. The minimum corresponds to the coarse-grained metastable equilibrium and the maximum to the unstable one.